\documentclass[11pt]{article}
\usepackage[margin=1in]{geometry}
\usepackage{graphicx}
\usepackage{url}
\usepackage{authblk}
\usepackage{amsmath}
\usepackage{enumitem}
\usepackage{natbib}
\usepackage{hyperref}
\hypersetup{colorlinks=true, linkcolor=blue, citecolor=blue, urlcolor=blue}

\title{A4L: An Architecture for AI-Augmented Learning}
\author[1]{Ashok Goel}
\author[1]{Ploy Thajchayapong}
\author[1]{Vrinda Nandan}
\author[1]{Harshvardhan Sikka}
\author[1]{Spencer Rugaber}
\affil[1]{Design Intelligence Laboratory, Georgia Institute of Technology}
\date{}

\begin{document}
\maketitle

\begin{abstract}
AI promises personalized learning and scalable education. As AI agents increasingly permeate education in support of teaching and learning, there is a critical and urgent need for data architectures for collecting and analyzing data on learning, and feeding the results back to teachers, learners, and the AI agents for personalization of learning at scale. At the National AI Institute for Adult Learning and Online Education, we are developing an Architecture for AI-Augmented Learning (A4L) for supporting adult learning through online education. We present the motivations, goals, requirements of the A4L architecture. We describe preliminary applications of A4L and discuss how it advances the goals of making learning more personalized and scalable.
\end{abstract}


\section{Introduction}
Artificial Intelligence (AI) has the potential to transform learning and education. On one hand, AI promises to personalize learning. We know that personalization of learning, for example through one-on-one tutoring, helps learners of all ages and at all levels of academic achievement. However, human instructors cannot do one-on-one tutoring for every student; AI already can do so for some classes of problems in some types of domains. On the other hand, AI promises to scale education. We know that online education can make education more accessible, affordable, and achievable, and thereby also more equitable. However, the quality of online education can vary a lot; AI already is helping enhance its quality especially for adult learners. The emergence of Generative AI has further accelerated these trends towards personalization at scale.

Personalization of learning however requires data about the learner and their learning. If a human instructor does not have data about a student’s prior knowledge and current learning experience, then it is difficult for the instructor to personalize the content, pace, or assessment of the student’s learning.  Similarly, if an AI agent does not have access to data about the learner and their learning, then, almost by definition, personalization is not feasible, no matter how powerful the agent might be in other dimensions. Fortunately, in modern ecosystems of education, learners and learning leave a rich digital footprint that can potentially be exploited to personalize learning \cite{unesco2023data}. Yet, current research on AI in learning and education focuses mostly on enhancing the power of AI agents without paying equal attention to collection, analysis and use of data on the learner and learning. We believe that there is a need to develop data architectures for collecting and analyzing data, and feeding the results back to instructors, learners, and the AI agents for personalization of learning at scale, while maintaining data privacy and security.

At the National Institute for AI Research for Adult Learning and Online Education (AI-ALOE) we are developing transformative AI techniques for enhancing the proficiency of online education for adult learning in STEM disciplines \cite{goel2024ai}. The current suite of AI technologies includes Apprentice Tutor for skill learning \cite{siddiqui2024htn}, SMART for concept learning \cite{kim2024summarization}, VERA for model learning \cite{an2021cognitive}, Jill Watson for conversational courseware \cite{taneja2024jwchatgpt}, Ivy for interactive videos \cite{dass2025enhanced}, and SAMI for social interactions \cite{kakar2024sami}. These AI tools have been extensively deployed and evaluated in a variety of classes at Georgia Tech as well as Technical College System of Georgia (TCSG).

Our goal at AI-ALOE is to study and develop educational ecosystems in which instructors, learners, and AI agents work together to enhance both learning and teaching. As part of this ecosystem, Georgia Tech’s Design Intelligence Laboratory is developing and implementing an \textit{Architecture for AI-Augmented Learning (A4L)} for collecting and analyzing data, and feeding the results back to the instructors, learners, and AI agents for personalization of learning at scale. Below, we present the motivations, goals, and requirements of A4L. We also describe preliminary applications of A4L and discuss how it advances the goals of making learning more personalized and scalable.

\section{Ecosystems for Adult Learning and Online Education}
A4L is grounded in ecosystems for adult learning and online education we are developing at AI-ALOE.  A key characteristic of adult learners ranging in age from, say, 20 years to 80 years old is that often they cannot leave their homes and workplaces for education due to factors such as health, family, finances, etc. Instead, we must take education to the working learners and learning workers. Online education offers a powerful medium for taking learning to adults where they live and work. A key characteristic of online education is its capacity to generate much richer data on learners and learning than traditional in-person classrooms. This is potentially useful for personalizing learning at scale if we can collect, store, analyze, and use the data. 
\subsection{An Ecosystem for Online Education}
Figure 1 illustrates elements of an “AI in the loop” ecosystem for online learning and teaching. Human instructors (shown in pale green at the top center of the figure) provide educational materials and instructions to learners (in grey at the center of the figure). The instructors may or may not take theories of human cognition and learning into account in their instruction; not all instructors do. Similarly, the instructors may or may not take the constraints and affordances of online learning environments into account; this is depicted through the dashed lines at the left center of the figure. 

We now introduce AI agents in the instructor-learner loop. AI agents (shown in pale purple in the bottom center of the figure) help teachers teach as well as learners learn. Unlike human instructors, we can ensure that these AI agents do take theories of cognition and learning as well as constraints and affordances of online environments into account. For example, AI agents may monitor the discussion forums of an online class, answer questions posted there and track the evolution of students’ questions.

Data about the student’s learning is collected continuously, not only through assessments but also through the student’s learning behaviors and interactions with the teacher, other students, and the AI agents. This rich data is mined, analyzed, and properly filtered for sharing with the teachers, the students, and the AI agents. The cycle repeats itself, leading to continuous, sustained, data-driven, evidence-based improvement in learning. This ecosystem for online education is aligned with the notion of learning engineering \cite{dede2018learning}. The A4L data architecture (not shown in Figure 1) is responsible for collecting, storing, anonymizing, standardizing, analyzing, visualizing, and sharing this data.

\begin{figure}[h]
\centering
\includegraphics[width=0.7\linewidth]{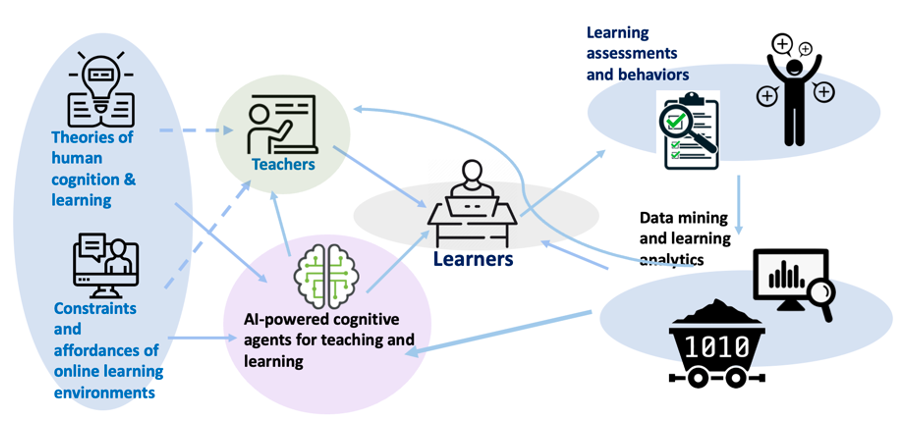}
\caption{\centering Elements of the AI-ALOE ecosystem for online learning and teaching. (The dashed arrows on the top left in figure indicate that the relationship is optional.)}
\end{figure}

\subsection{An Ecosystem for Adult Learning}
\cite{lyndgaard2024lifelong} at AI-ALOE describe an ecosystem for adult learning. Their framework encompasses micro-learning over (very) short periods of time (for example, cognition and meta-cognitive processes of learning, managing how to learn), meso-learning over intermediate periods of time (social and emotional aspects of learning, managing motivation to learn over time), and macro-learning over (very) long periods of time (managing time to learn given competing demands from family, work, etc.). 
\begin{figure}[h]
\centering
\includegraphics[width=0.7\linewidth]{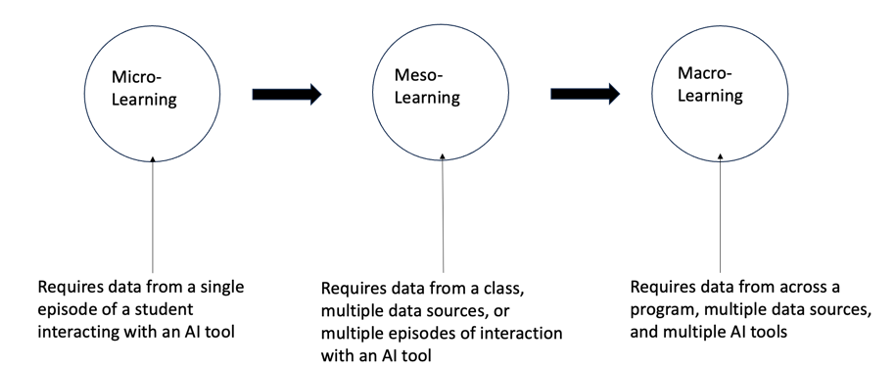}
\caption{\centering Translating Micro-, Meso-, and Macro-Learning into data requirements.}
\end{figure}

Figure 2 translates Lyndgaard, Storey, \& Kanfer’s ecosystem for adult learning into data requirements for A4L in the context of the AI-in-the-loop ecosystem for online education presented above. Micro-learning requires data from individual episodes of a student interacting with an AI tool such as Jill Watson. Meso-learning requires data from multiple episodes of interaction with an AI tool, or data from multiple data sources including, for example a Learning Management System, or data from a whole class both in terms of a student’s trajectory in a class and data of other students in the class. Macro-learning requires data from multiple AI tools, and multiple data sources, and an educational program (or a series of programs). A significant portion of AI research on learning addresses micro-learning. A4L seeks to address both micro- and meso-learning (and potentially macro-learning as well).

\subsection{Data Requirements for Online Education for Adult Learners}
Data is crucial in AI-enhanced online education to enabling personalized learning at scale. By analyzing student behaviors, engagement, and progress, teachers and AI agents alike can refine teaching strategies for more effective and inclusive learning. As indicated above; to maximize AI-driven education, data must be collected at three levels:

\paragraph{Micro-learning (Individual Learning Episodes):}       
This level focuses on short, focused activities. Example data include:
\begin{itemize}[noitemsep]
    \item \textit{Clickstream data:} Tracking student interactions with AI tools.
    \item \textit{Engagement metrics:} Measuring attention, persistence, and interaction frequency.
    \item \textit{Cognitive patterns:} Identifying problem-solving strategies.
\end{itemize}

\paragraph{Meso-learning (Learning Over Time):}
This level captures trends over a semester or course. Example data include:
\begin{itemize}[noitemsep]
    \item \textit{Student trajectories:} Monitoring performance trends.
    \item \textit{Collaboration data:} Tracking discussions and group work.
    \item \textit{Learning Management System data:} Analyzing assignments, quizzes, and feedback.
    \item \textit{Motivational data:} Understanding factors like self-efficacy and peer learning.
\end{itemize}

\paragraph{Macro-learning (Long-Term Educational Progress):}
This level tracks lifelong learning patterns. Example data include:
\begin{itemize}[noitemsep]
    \item \textit{Cross-course performance:} Aggregating long-term trends.
    \item \textit{Multi-source integration:} Merging data from LMS, AI tools, and institutional records.
\end{itemize}

\section{The Architecture for AI-Augmented Learning (A4L)}
The A4L architecture is specifically designed to meet the diverse data needs of the AI-ALOE ecosystem for online education and adult learning. It offers a robust suite of features to streamline data processing and improve data quality. This section outlines the \textit{conceptual architecture} of A4L data pipeline, as illustrated in Figure 3, with a focus on its key capabilities: data upload, data standardization, data anonymization, data storage, and data deanonymization and data download. 
\begin{figure}[h]
\centering
\includegraphics[width=0.7\linewidth]{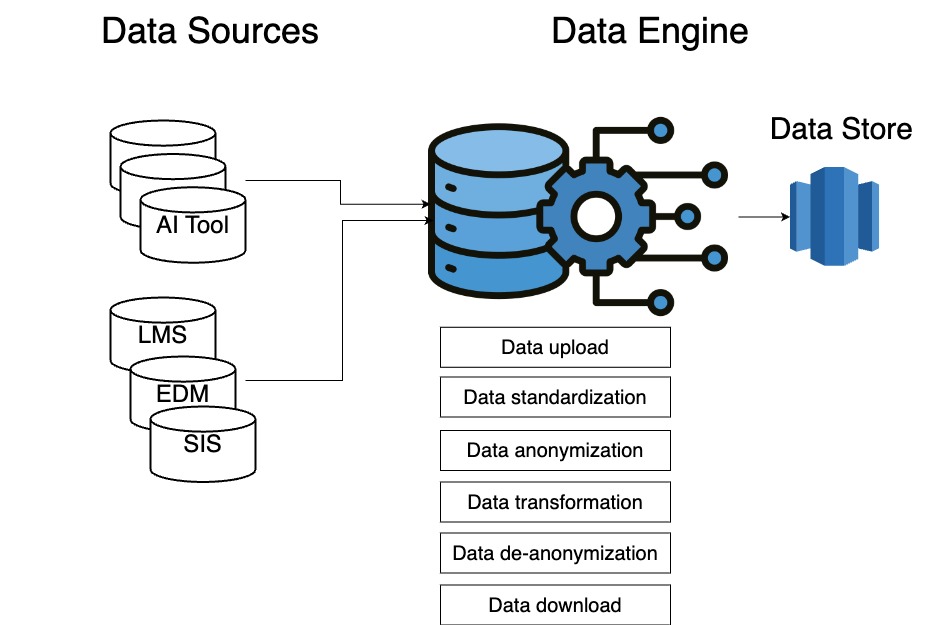}
\caption{\centering The A4L Data Pipeline Conceptual Architecture.}
\end{figure}

\subsection{Data Sources}
The A4L Data Pipeline collects data from various sources, including Learning Management Systems (LMS), class discussion forums, and AI tools such as Jill Watson. The data is standardized to 1EdTech’s Caliper Analytics \footnote{\url{https://www.1edtech.org/standards/caliper}} standards and downloaded when needed. Additionally, data from Enterprise Data Management (EDM) plaforms and Student Information Systems (SIS) can be manually uploaded through cloud data repositories, offering flexibility in data ingestion. This design enables the pipeline to process both structured and unstructured data ensuring compatibility with diverse data formats.

\subsection{Data Engine}
A4L Data Engine is central to the feedback loop in our ecosystem for online education (Figure 1), facilitating data flow and processing from different sources as mentioned in Section 3a. The data, which spans various formats and structures—including structured tables from LMS, clickstream data, and unstructured text from discussion forums—poses a significant challenge: This data contains Personally Identifiable Information (PII), necessitating stringent privacy measures and innovative techniques to identify, separate, and protect PII, particularly within the unstructured data.

In addition to managing a wide range of data, the primary function of the A4L Data Engine is to process and make this data accessible to downstream services. These include advanced analytics, intuitive visualizations and dashboards that empower educators and learners to gain insights, make informed decisions, and ultimately improve learning outcomes. A key differentiator of the A4L data pipeline, setting A4L apart from similar initiatives, is its ability to support and enable systematic personalization of AI technologies such as Apprentice Tutor and SMART mentioned earlier.  

The A4L Data Engine includes the following components, each designed for scalability to efficiently handle growing data volumes and increasing data variety and complexity:

\subsubsection{Data Upload and Standardization }
Data from multiple sources can be automatically or manually uploaded. The system is built to scale seamlessly, ensuring that as new data sources are added or as data volume increases, the upload process remains efficient. Modules ensure the data conforms to Caliper standards, creating a consistent structure that allows seamless processing in downstream stages, regardless of the amount of incoming data.

\subsubsection{Data Anonymization }
Anonymization is crucial to protect PII across both structured and unstructured data. The anonymization process can be scaled to accommodate large datasets, utilizing advanced techniques like PIILO, a large language model specifically designed to detect and obfuscate PII within unstructured text data \cite{holmes2023piilo}. For structured data, obfuscated identities replace student identifiers, maintaining data relationships without exposing sensitive information. This scalable approach ensures privacy protection as data grows in volume and complexity.

\subsubsection{Data Transformation }
Once cleansed, data undergoes transformation, converting disparate data into structured tables. The transformation process is highly scalable, capable of handling large and diverse datasets as the system expands. This stage ensures data consistency, quality, and readiness for analytics or further applications, while supporting increasingly complex datasets and larger-scale operations.

\subsubsection{Data Store }
The final stage of the processing in the A4L Data Engine is the consolidation of processed data in a centralized, secure data store. This storage solution is designed to scale with increasing data volumes, offering efficient storage and rapid retrieval. The centralized repository is accessible through data download application, allowing flexible and scalable access for users and the creation of visualizations and downstream applications. This scalability ensures that as data usage grows, the system can continue to provide fast and reliable access for uses like model training and analytics. The data store is optimized for integration within the AI-ALOE ecosystems for online education for adult learners, supporting scalability and personalization at scale.

\subsection{Data Model}
\begin{figure}[h]
\centering
\includegraphics[width=0.7\linewidth]{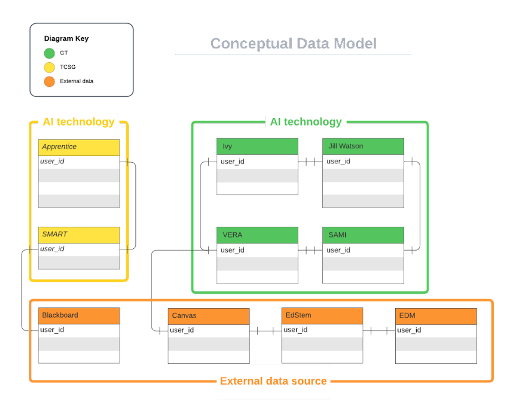}
\caption{\centering Conceptual Data Model for linking student data across AI tools. (The AI agents in green have been deployed at Georgia Tech (GT) while those in yellow have been deployed at the Technical College System of Georgia (TCSG).)}
\end{figure}
The Data Model maps connections between data sources, detailing data types, relationships, and structure. Designed at varying abstraction levels, it aligns with research goals involving AI tools like Apprentice Tutor, Jill Watson, SAMI, SMART, and VERA. Student data from LMS systems—such as enrollment and demographics—is only processed after anonymization to ensure privacy.

The model links student data with AI tool outputs, supporting a shift from micro- to meso-learning research. As shown in Figure 4, yellow and green blocks represent micro-learning data from AI tools, while orange blocks show meso-level connections to external sources. Identity mapping is maintained separately for Georgia Tech and TCSG.

Structured to support analytics, the data model enhances data-driven personalization and learning within the AI-ALOE ecosystem and enables insight generation through: 

\begin{itemize}[noitemsep]
    \item Defining data needs and research goals
    \item Gathering accurate and current data
    \item Integrating data via A4L (cleaning, transforming, standardizing)
    \item Storing data in structured databases for efficient analysis
\end{itemize}

As Figure 5 illustrates, we have extended the A4L data pipeline to include Data Analytics and Visualization modules. We describe these modules next.
\begin{figure}[h!]
\centering
\includegraphics[width=0.7\linewidth]{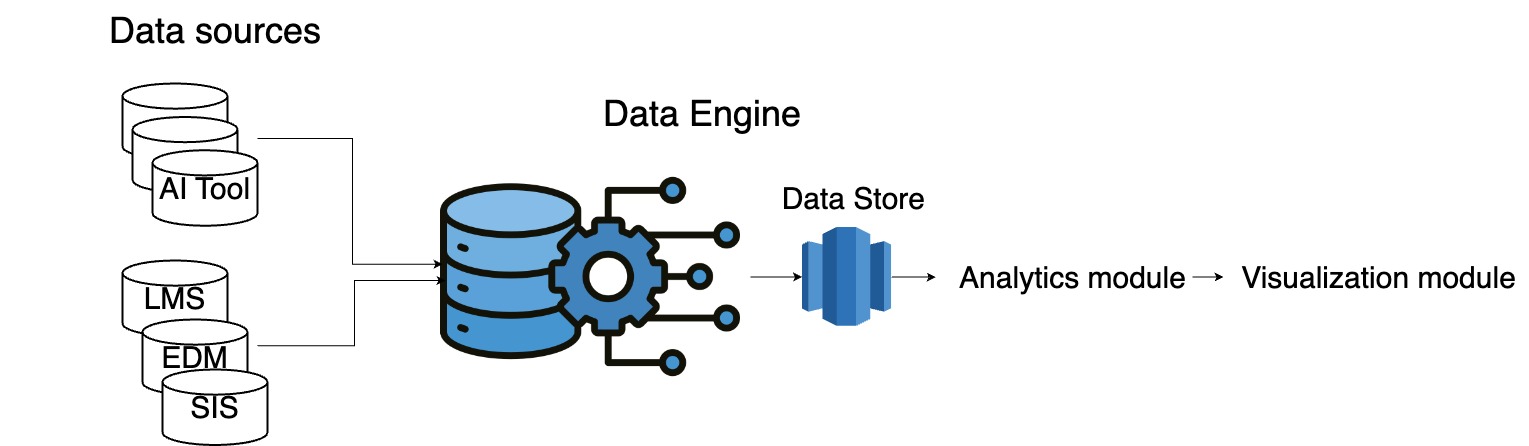}
\caption{\centering Extensions to the A4L Data Pipeline include Analytics and Visualization modules on the right}
\end{figure}

\subsection{Analytics Module}
Data is transferred from the data store into the Analytics module through a time-based scheduled job. 

\paragraph{Data Flow:}
\begin{itemize}[noitemsep]
    \item \textit{Time-Based Scheduled Job:} The process begins with a scheduled job that triggers data processing at predefined intervals. This ensures regular updates and analysis of incoming data.
    \item \textit{Data Analysis Module:} The data is then processed through an analysis module that extracts insights from the collected information. This module may include various analytical techniques such as statistical analysis and machine learning.
    \item \textit{Analysis Results:} After processing, the module generates results that can be used for visualization and reporting. 
\end{itemize}
This Data Analytics pipeline ensures timely and automated data processing, enabling efficient analytics and insights while providing the flexibility to expand and scale as needed. We will discuss in further details how the Analytics pipeline was implemented based on the application from the AI tools in Section 4.

\subsection{Visualization Module}
The results from the Analytics module are sent to the Visualization Module (Figure 6), where it is transformed into visually informative and actionable representations. This enriched data becomes available to users (teachers, learners, researchers) through a dashboard, providing a user-friendly interface for accessing and interacting with the data.
\begin{figure}[h]
\centering
\includegraphics[width=0.7\linewidth]{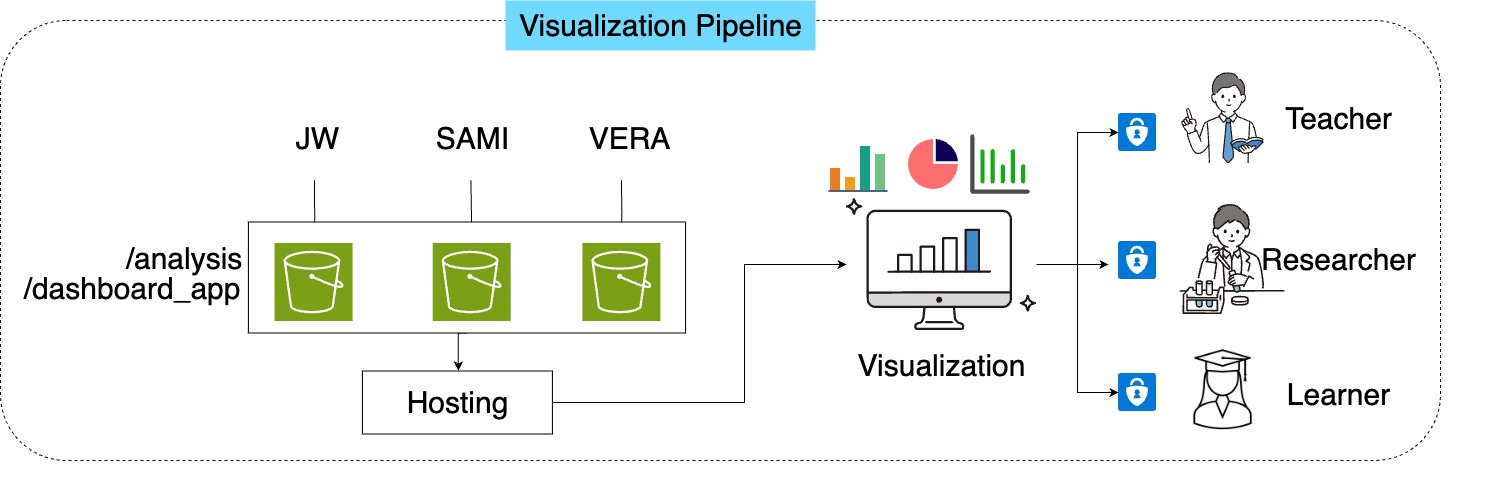}
\caption{High-level Architecture of the Visualization Module}
\end{figure}
\paragraph{Data Flow:}
\begin{itemize}[noitemsep]
    \item \textit{Visualization Generation:} The data is processed and visualized using three different tools: Jill Watson, SAMI, and VERA. Each tool represents different visualization libraries or frameworks.
    \item \textit{Visualization Hosting:} The generated visualizations are hosted on platform and served to the end-users.
    \item \textit {User Access:} The visualizations are accessed by three user groups: Teachers, Researchers, and Learners. Each group has different access permissions and views into the visualizations.
\end{itemize}

\section{Preliminary Applications of A4L}
In this section, we briefly describe three preliminary applications of the A4L data architecture in the context of the ecosystem of online education for adult learning described previously in Section 2. Specifically, we are presently using the A4L Analytics Pipeline mentioned in Section 3d to analyze data from three different AI technologies: VERA for model learning, Jill Watson for conversational courseware, and SAMI for social interactions. All three applications support both micro- and meso-learning,  illustrating the potential of A4L to support personalization of learning in online education for adult learners.

\subsection{Illustrative Example 1: VERA}
VERA is an online interactive learning tool that allows users to construct conceptual models of complex systems and validate them through simulation \cite{an2021cognitive}. Its primary function is to support inquiry-based learning (IBL) in ecological modeling. For instance, users can construct predator-prey models to explore how population dynamics fluctuate over time based on different species' characteristics. 

In the context of the ecosystem for adult learning described in Section 2b, VERA entails both micro-learning and meso-learning. \cite{an2021cognitive} examined students' cognitive strategies for estimating values for the parameters of an ecological model. The findings revealed three distinct parameter estimation strategies: systematic search, problem decomposition/reduction, and global/local search. These patterns align with micro-learning, where data is derived from individual episodes of interaction with VERA. 

As part of meso-learning, \cite{fryer2025ai} integrated survey data from a class where VERA was used to explore how cognitive traits, motivational factors, and learning strategies influence tool adoption. They analyzed need for cognition (NFC), self-efficacy (SE), help-seeking (HS), and peer learning (PL) among students. Results showed that adopters of VERA had significantly higher NFC, suggesting that students who prefer deeper cognitive engagement are more inclined to use AI tools. However, these adopters had lower HS and PL scores, indicating a potential preference for human-based assistance over AI support, highlighting some limitations in tool adoption. This analysis has been successfully implemented within the A4L Analytics Pipeline, providing a structured approach to examining learner interactions, cognitive behaviors, and AI tool adoption trends in educational settings.

\subsection{Illustrative Example 2: Jill Watson}
Jill Watson (JW) is a conversational AI teaching assistant designed to support student learning by answering questions and fostering discussion based on course materials \cite{goel2018jill} and \cite{taneja2024jwchatgpt}. It is hosted on 1EdTech’s  LTI \footnote{\url{https://www.1edtech.org/standards/lti}} standard for LMS. While AI tools like JW can enhance education, their adoption and effectiveness may vary across demographics, raising equity concerns. To explore this, we studied how demographic factors affect AI adoption and student performance among adult learners \cite{fryer2025ai}. Results showed a significant difference in adoption rates between Gen-Z and Pre-Gen Z students. This study is now integrated into the A4L Analytics Pipeline. We are expanding the pipeline to include analyses from \cite{maiti2025jwquestions}, who studied over 5,500 students across online and hybrid classes. They used fine-tuned BERT and regression models to classify student questions to JW by cognitive complexity. Findings showed an increase in higher-order questions over time, suggesting JW promotes critical thinking. 

These studies include both micro- and meso-learning analyses. Micro-learning focuses on individual interactions with Jill Watson, assessing immediate engagement and learning. Over time, these interactions build toward deeper inquiry. Meso-learning analyzes patterns across multiple students and systems (LMS, SIS), offering insights into broader learning trajectories and class-wide outcomes. Results showed older students were more likely to adopt Jill Watson. Access to Jill Watson correlated with better academic performance. Together, micro- and meso-level insights support a more personalized and effective learning experience through AI.

\subsection{Illustrative Example 3: SAMI}
SAMI is an AI-powered social assistant designed to help students in large online classes connect with peers who share similar interests and traits \cite{kakar2024sami}. Its goal is to boost social presence in asynchronous learning environments and reduce the negative effects of isolation on students’ well-being and learning experiences. 

\cite{wojcik2025samiimpact} evaluated SAMI’s impact by examining student participation, engagement, introductions, connection success, demographics, personality traits, and academic performance. Early results show that SAMI users engage more—asking and answering questions, commenting, and receiving likes—suggesting greater social involvement. The study also explored motivations for using SAMI through Need for Belonging (NTB), Self-Efficacy (SE), and personality traits. Students using SAMI reported a higher sense of belonging. NTB also correlated with traits like extraversion and neuroticism, highlighting the role of social needs in AI adoption.

SAMI operates within micro- and meso-learning frameworks. Micro-learning captures individual interactions and social behavior, while meso-learning tracks trends across time and the class. This dual approach helps SAMI support both personal and class-wide social engagement. SAMI is now integrated into the A4L Analytics Pipeline.

\subsection{Bi-Directional Feedback Loops}
The VERA, Jill Watson, and SAMI examples above illustrate the usefulness of A4L to support both micro- and meso-learning. A4L is designed to support exchange of signals between instructors and learners that plays a crucial role in effective instruction. It allows both parties to respond and adapt based on each other’s behaviors. This form of bidirectional feedback ensures that teaching is dynamically shaped by the learner’s behaviors and the teacher’s awareness of those behaviors. In this context, personalization is the mechanism through which bi-directional feedback becomes actionable at scale \cite{thajchayapong2025aipipeline}. For learners, personalization means that instructors and AI agents can respond to their individual needs, misconceptions, and preferences during learning.  For instructors, it means building insights into at both individual and classroom level through individual and aggregated student data, and tailor instruction accordingly. These feedback loops help teachers refine instruction, support learners more effectively, and even learn from the process themselves.
\begin{figure}[h]
\centering
\includegraphics[width=0.7\linewidth]{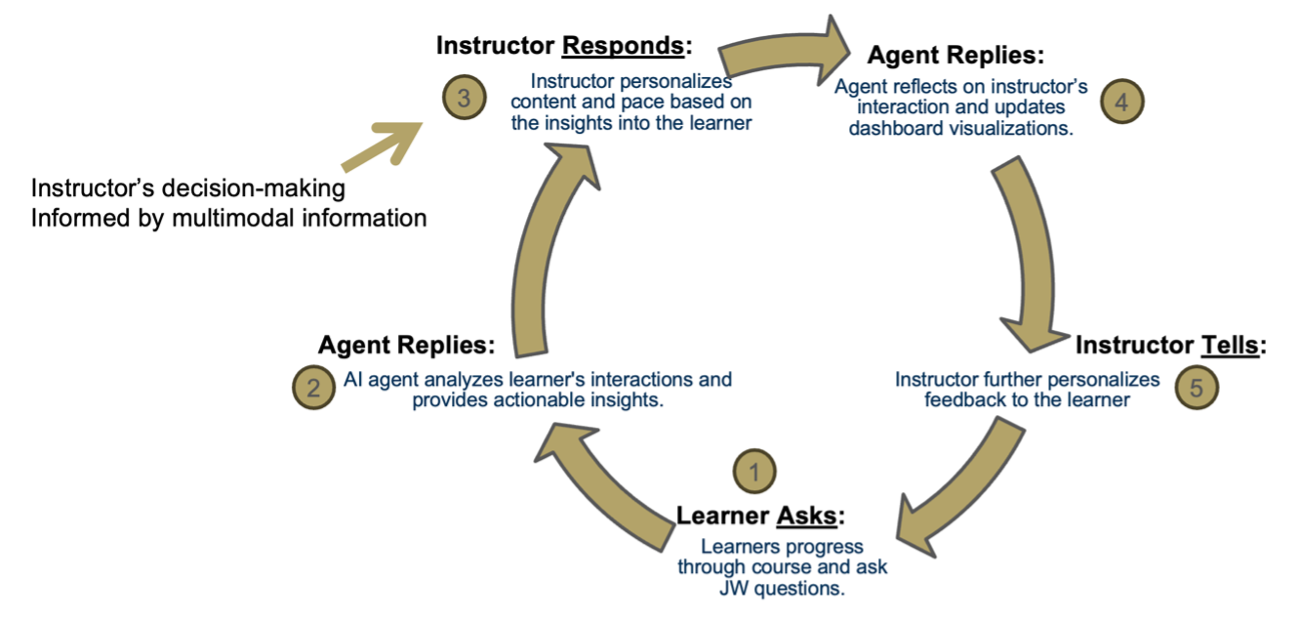}
\caption{\centering The instructor-AI-Learner information feedback loop. Learners (at the base of the figure) interact with Jill Watson (JW), generating data on their learning.  The A4L pipeline transforms this data into visualizations.  Instructors use these insights to personalize their instruction. (Adapted from \cite{thajchayapong2025aipipeline})}
\end{figure}

Figure 7 illustrates an example of the instructor-AI-learner feedback loop. The process follows a cycle of \textbf{Ask → Respond → Tell}, representing a bi-directional feedback loop between instructors and learners mediated by Jill Watson. \textbf{Ask}: As learners engage with the class materials, they complete assigned tasks and engage in conversations with Jill Watson. These interactions generate data about the student behavior and needs. \textbf{Respond}: A4L collects the above data, analyzes it, and visualizes it in a dashboard. Instructors review the dashboard and make informed decisions, adjusting their instructional focus or modifying course contents based on the insights from learner-AI interactions. \textbf{Tell}: Jill Watson reflects on the instructor’s feedback to the learner and updates the visualizations on the dashboard. The instructor provides additional personalized feedback based on the new insights.

\section{Related Data Architectures}
A4L stands apart from other data architectures by serving as a tailored data pipeline for AI-ALOE’s ecosystem for online education for adult learners, focusing on AI-augmented personalization of learning at scale. 

\subsection{Total Learning Architecture (TLA)}
Advanced Distributed Learning (ADL) Initiative’s Total Learning Architecture (TLA) is a robust data architecture used widely in the U.S. Department of Defense (Schatz \& Vogel-Walcutt). It supports interoperable learning technologies using the xAPI standard. A4L draws inspiration from TLA but emphasizes AI-augmented personalization for adult online learners, rather than general-purpose integration.

\subsection{Unizin}
Unizin \footnote{\url{https://unizin.org/knowledge-base/}} is a higher education consortium offering a cloud-based platform for student data analytics and decision-making. While Unizin supports various LMSs and educational tools, A4L is specifically designed to enable AI-augmented personalization at scale.

\subsection{My Learning Analytics}
My Learning Analytics \footnote{\url{https://its.umich.edu/academics-research/teaching-learning/my-learning-analytics/}} (MyLA) provides visual dashboards in Canvas to help track student engagement and performance. In contrast, A4L is a broader architecture that supports data-driven AI-augmented personalization of learning.

\subsection{DataShop and LearnSphere }
DataShop and LearnSphere platforms host K-12 learning data for research and instructional improvement (Koedinger et al. 2010). A4L too acts like a data repository focusing on adult online education and supports AI-ALOE AI tools deployed at Georgia Tech and TCSG.

\subsection{Data for AI Research }
The Data for AI Research supports integrated data lakes and real-time adaptive feedback \cite{joksimovic2022dair}. A4L shares these features but is tailored to the data needs AI-ALOE ecosystems for AI-augmented adult learning in online education. 

\section{Conclusion}
In this chapter, we presented an ecosystem for online education based on information feedback loops that support continuous and sustained improvement in learning. We also presented an ecosystem for adult learning encompassing micro-, meso- and macro-learning. We then sketched the outlines of A4L, a data architecture for AI-augmented learning in the ecosystem for online education for adult learners. While some parts of A4L are presently under development, others already have been deployed for preliminary AI applications. We described how A4L supports meso-learning in Jill Watson that aids cognitive engagement through conversational courseware, in SAMI that enhances social interactions, and in VERA that helps improve model learning.

Our experiences with the use of A4L for supporting Jill Watson, SAMI, and VERA in various learning contexts have led to a deeper understanding of the design requirements for a data architecture that can support personalization at scale. These design requirements form the basis for A4L2.0 that we are developing in collaboration with 1EdTech. Among many other features, A4L2.0 provides a data pipeline for automatic ingestion of LMS, EDM, and SIS data as well as standardization of data in Caliper Analytics. We expect that this systemization of data ingestion and standardization of the data store will enable sharing of data with the larger learning and education research community while preserving data privacy and security. As we push A4L2.0 towards supporting macro-learning, we expect that it will include data not only on learning but education in general, for example, data on admissions, financial aid, extracurricular activities, and educational history.

\section*{Acknowledgements}
This research has been supported by grants from NSF grants US National Science \#2247790 and \#2112532 to the National AI Institute for Adult Learning and Online Education (aialoe.org). We thank members of the A4L team for their contributions to this work including our 1EdTech colleagues Suzanne Carbonaro, Blaine Helmick, Tim Couper, and Andrea Deau.

\bibliographystyle{plainnat}
\bibliography{a4l_refs}

\end{document}